\documentclass[aps,prl,reprint,superscriptaddress,amsmath]{revtex4-1}
\usepackage{graphicx}
\usepackage{bbold}
\usepackage{hyperref}
\usepackage{geometry}
\usepackage{amsmath}

\newcommand{\bestfidelity}{97.4(2)~\%} 
\newcommand{\meanfidelity}{96.9(1)~\%} 
\newcommand{\concurrence}{95.2(5)~\%} 
\newcommand{\chsh}{2.75(1)} 
\newcommand{\efficiency}{5.7\%} 
\newcommand{\correlations}{4722} 
\newcommand{\rate}{40.5} 
\newcommand{\bra}[1]{\ensuremath{\langle #1|\,}}
\newcommand{\ket}[1]{\ensuremath{\,|#1\rangle}}

\newcounter{firstbib}

\begin{document}


\title{Tunable Ion-Photon Entanglement in an Optical Cavity}



\author{A. Stute}
\affiliation{Institut f{\"u}r Experimentalphysik, Universit{\"a}t Innsbruck, Technikerstra{\ss}e 25, 6020 Innsbruck, Austria}
\author{B. Casabone}
\affiliation{Institut f{\"u}r Experimentalphysik, Universit{\"a}t Innsbruck, Technikerstra{\ss}e 25, 6020 Innsbruck, Austria}
\author{P. Schindler}
\affiliation{Institut f{\"u}r Experimentalphysik, Universit{\"a}t Innsbruck, Technikerstra{\ss}e 25, 6020 Innsbruck, Austria}
\author{T. Monz}
\affiliation{Institut f{\"u}r Experimentalphysik, Universit{\"a}t Innsbruck, Technikerstra{\ss}e 25, 6020 Innsbruck, Austria}
\author{P.~O.~Schmidt}
\affiliation{QUEST Institute for Experimental Quantum Metrology, Physikalisch-Technische Bundesanstalt, 38116 Braunschweig, Germany}
\affiliation{Institut f\"ur Quantenoptik, Leibniz Universit\"at Hannover, 30167 Hannover, Germany}
\author{B. Brandst\"atter}
\affiliation{Institut f{\"u}r Experimentalphysik, Universit{\"a}t Innsbruck, Technikerstra{\ss}e 25, 6020 Innsbruck, Austria}
\author{T.~E. Northup}
\affiliation{Institut f{\"u}r Experimentalphysik, Universit{\"a}t Innsbruck, Technikerstra{\ss}e 25, 6020 Innsbruck, Austria}
\author{R. Blatt}
\affiliation{Institut f{\"u}r Experimentalphysik, Universit{\"a}t Innsbruck, Technikerstra{\ss}e 25, 6020 Innsbruck, Austria}
\affiliation{\"Osterreichischen Akademie der Wissenschaften, Technikerstra{\ss}e 21a, 6020 Innsbruck, Austria}


\date{\today}

\begin{abstract}
Publication reference: Nature \textbf{485}, 482--485  (2012), \href{http://www.nature.com/nature/journal/v485/n7399/full/nature11120.html}{doi:10.1038/nature11120} \\ \\
\textbf{
Proposed quantum networks require both a quantum interface between light and matter and the coherent control of quantum states \cite{Cirac97,Kimble08a}.
A quantum interface can be realized by entangling the state of a single photon with the state of an atomic or solid-state quantum memory, as demonstrated in recent experiments with trapped ions \cite{Blinov04, Olmschenk09}, neutral atoms \cite{Volz06,Wilk07b}, atomic ensembles \cite{Matsukevich05,Sherson06}, and nitrogen-vacancy spins \cite{Togan10}.
The entangling interaction couples an initial quantum memory state to two possible light-matter states, and the atomic level structure of the memory determines the available coupling paths. In previous work, these paths' transition parameters determine the phase and amplitude of the final entangled state, unless the memory is initially prepared in a superposition state \cite{Olmschenk09}, a step that requires coherent control.
Here we report the fully tunable entanglement of a single $\mathbf{^{40}\mathrm{\textbf{Ca}}^{+}}$ ion and the polarization state of a single photon within an optical resonator.
Our method, based on a bichromatic, cavity-mediated Raman transition, allows us to select two coupling paths and adjust their relative phase and amplitude.
The cavity setting enables intrinsically deterministic, high-fidelity generation of any two-qubit entangled state.
This approach is applicable to a broad range of candidate systems and thus presents itself as a promising method for distributing information within quantum networks.}
\end{abstract}

\pacs{}

\maketitle 



%
%

%

Optical cavities are often proposed as a means to improve the efficiency of atom-photon entanglement generation.  Experiments using single emitters \cite{Blinov04,Volz06,Olmschenk09,Togan10} collect photons over a limited solid angle, with only a small fraction of entanglement events detected.  However, by placing the emitter inside a low-loss cavity, it is possible to generate photons with near-unit efficiency in the cavity mode \cite{Law97,Cirac97}.
Neutral atoms in a resonator have been used to generate polarization-entangled photon pairs \cite{Wilk07b, Weber09},
but this has not yet been combined with coherent operations on the atomic state.
Trapped ions have the advantage of well-developed methods for coherent state manipulation and readout \cite{Leibfried03,Haeffner08}.  Using a single trapped ion integrated with a high-finesse cavity, we
implement full tomography of the joint atom-photon state and
generate maximally entangled states with fidelities up to \bestfidelity.

In initial demonstrations of atom-photon entanglement, the amplitudes of the resulting state are fixed by atomic transition amplitudes \cite{Blinov04, Volz06, Wilk07b, Weber09, Togan10}. If the final atomic states are not degenerate, as in the case of a Zeeman splitting, the phase of the atomic state after photon detection is determined by the time at which detection occurs.
In contrast, we control both amplitude and phase via two simultaneous cavity-mediated Raman transitions. The bichromatic Raman fields ensure the independence of the atomic state from the photon-detection time; their relative amplitude and phase determine the state parameters.
Within a quantum network, such a tunable state could be matched to any second state at a remote node,
generating optimal long-distance entanglement in a quantum-repeater architecture \cite{Briegel98}.

A tunable state has previously been employed as the building block for teleportation \cite{Olmschenk09} and a heralded gate between remote qubits \cite{Maunz09}. In this case, tunability of the entangled state is inherited from control over the initial state of the atom. 
The photonic qubit is encoded in frequency, and as a result, integration with a cavity would be technically challenging.
The entangling process is intrinsically probabilistic, with efficiency limited to 50\% even if all emitted photons could be collected.
In the scheme presented here, the entangling interaction itself is tunable, and no coherent manipulation of the input state is required.
For atomic systems with a complex level scheme in which several transition paths are possible, the two most suitable paths can be selected.

\begin{figure}
\includegraphics[width = \columnwidth]{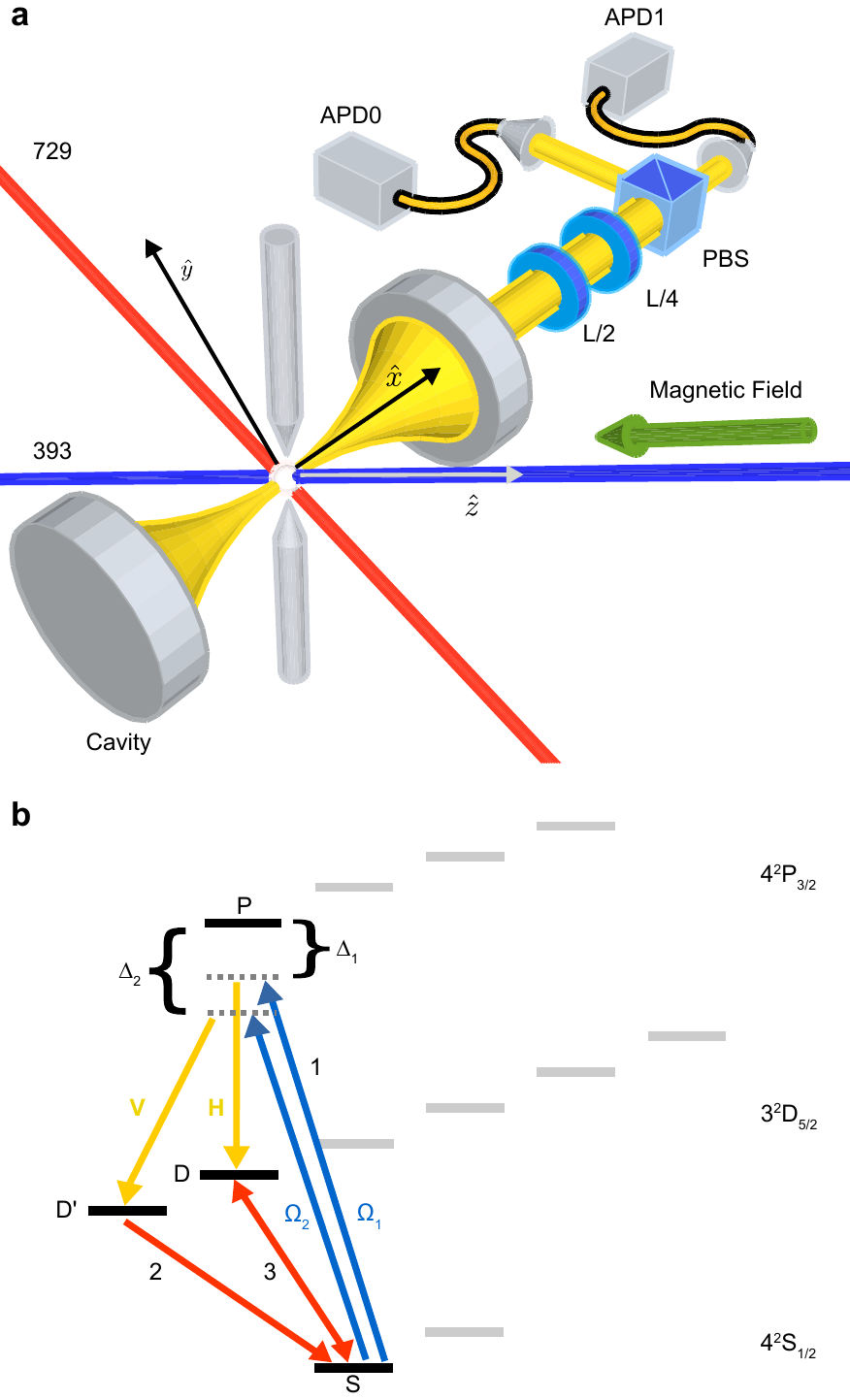}
\caption{\label{setup} \textbf{Experimental apparatus and entanglement sequence.} \textbf{a}, An ion is confined in a Paul trap (indicated by two endcaps) at the point of maximum coupling to a high-finesse cavity. A 393-nm laser generates atom-photon entanglement, characterized using a 729-nm laser. Photons' polarization exiting the cavity is analyzed using half- and quarter-waveplates (L/2, L/4), a polarizing beamsplitter cube (PBS), and fiber-coupled avalanche photodiodes (APD0, APD1).  \textbf{b}, A bichromatic Raman pulse with Rabi frequencies $\Omega_{1}, \Omega_{2}$ and detunings $\Delta_1, \Delta_2$ couples $| S \rangle$ to states $| D \rangle$ and $| D' \rangle$ via two cavity modes $H$ and $V$ (1), generating a single cavity photon.  To read out entanglement, $| D' \rangle$ is mapped to $| S \rangle$ (2), and coherent operations on the $S-D$ transition (3) prepare the ion for measurement.}
\end{figure}

Our experimental apparatus (Fig. \ref{setup}(a)) consists of a linear Paul trap storing a single  $^{40}\mathrm{Ca}^{+}$ ion within a 2~cm optical cavity \cite{Russo09,Stute11}.
The cavity has a waist of 13~$\mu$m and finesse  of 77,000 at 854~nm, the wavelength of the $4^{2}P_{3/2}-3^{2}D_{5/2}$ transition. The rates of coherent atom-cavity coupling $g$, cavity-field decay $\kappa$, and atomic polarization decay $\gamma$ are given by ($g, \kappa,\gamma) = 2\pi \times (1.4, 0.05, 11.2)$~MHz.
The ion is located in both the waist and in an antinode of the cavity standing wave, and it is localized to within $13\pm7$~nm along the cavity axis \cite{Stute11}.
Entanglement is generated via a bichromatic Raman field at 393~nm and read out using a quadrupole field at 729~nm.

A magnetic field of 2.96~G is applied along the quantization axis $\hat{z}$ and perpendicular to the cavity axis.
The cavity supports degenerate horizontal ($H$) and vertical ($V$) polarization modes, where $H$ is defined parallel to $\hat{z}$. At the cavity output, the modes are separated on a polarizing beamsplitter and detected at avalanche photodiodes.  A half- and a quarter-waveplate prior to the beamsplitter allow us to set the measurement basis of the photon \cite{James01}.

The entangling process is illustrated in Fig. \ref{setup}(b). Following a Doppler-cooling interval, the ion is initialized via optical pumping in the state $| S \rangle \equiv | 4^{2}S_{1/2}, m_{S} = -1/2 \rangle$.
In order to couple $| S \rangle$ simultaneously to the two states $| D \rangle \equiv |3^{2}D_{5/2}, m_{D} = -3/2 \rangle$ and $| D' \rangle \equiv |3^{2}D_{5/2}, m_{D} = -5/2 \rangle$, we apply a phase-stable bichromatic Raman field, detuned by $\Delta_1$ and $\Delta_2$ from the $\ket{S}-\ket{P}$ transition. Here, the intermediate state $\ket{P} \equiv \ket{4^{2}P_{3/2},m_P=-3/2}$ is used.
The cavity is stabilized at detuning $\Delta^{\mathrm{c}}_1 \approx -400$~MHz from the $\ket{P}-|D\rangle$ transition and $\Delta^{\mathrm{c}}_2 = \Delta^{\mathrm{c}}_1 + \Delta_{D,D'}$ from the $\ket{P}-|D'\rangle$ transition, where $\Delta_{D,D'}$ is the Zeeman splitting between $\ket{D}$ and $\ket{D'}$.
When $\Delta^{\mathrm{c}}_i$ and $\Delta_i$
satisfy the Raman resonance condition for both $i=(1,2)$, population is transferred coherently from $\ket{S}$ to both $\ket{D}$ and $\ket{D'}$, and a single photon is generated in the cavity \cite{McKeever04a,Keller04,Hijlkema07,Barros09}.

The effective coupling strength of each of the two transitions is given by $g^{\mathrm{eff}}_i = \Omega_i G_i g /\Delta_i$. Here, $\Omega_{1}$ and $\Omega_{2}$ are the amplitudes of the Raman fields;
$G_1$ and $G_2$ are the products of the Clebsch-Gordon coefficients and the projections of laser and vacuum-mode polarizations onto the atomic dipole moment \cite{Stute11}.
In free space, these two pathways generate $\pi$- and $\sigma^{+}$-polarized photons, respectively.  Within the cavity, the $\pi$ photon is projected onto $H$ and the $\sigma^{+}$ photon onto $V$ \cite{Russo09,Stute11}.
Ideally, the bichromatic fields generate any state of the form
\[
|\psi \rangle = \cos{\alpha} | D H \rangle + e^{i\varphi} \sin{\alpha}| D'  V \rangle,
\]
where $\alpha \equiv \tan^{-1}{(g^{\mathrm{eff}}_2/g^{\mathrm{eff}}_1)}$ and $\varphi$ is determined by the relative phase of the Raman fields.
To determine the the overlap of the measured state with $\ket{\psi}$, we perform quantum state tomography of the ion-photon density matrix $\rho$ for given values of $\alpha$ and $\varphi$.
Ion and photon are measured in all nine combinations of ion Pauli bases $\{ \sigma_{x}, \sigma_{y}, \sigma_{z} \}$ and photon polarization bases \\ $\{ H/V, \mathrm{diagonal/antidiagonal, right/left} \}$ \cite{James01}.

\begin{figure}
\includegraphics[width=\columnwidth]{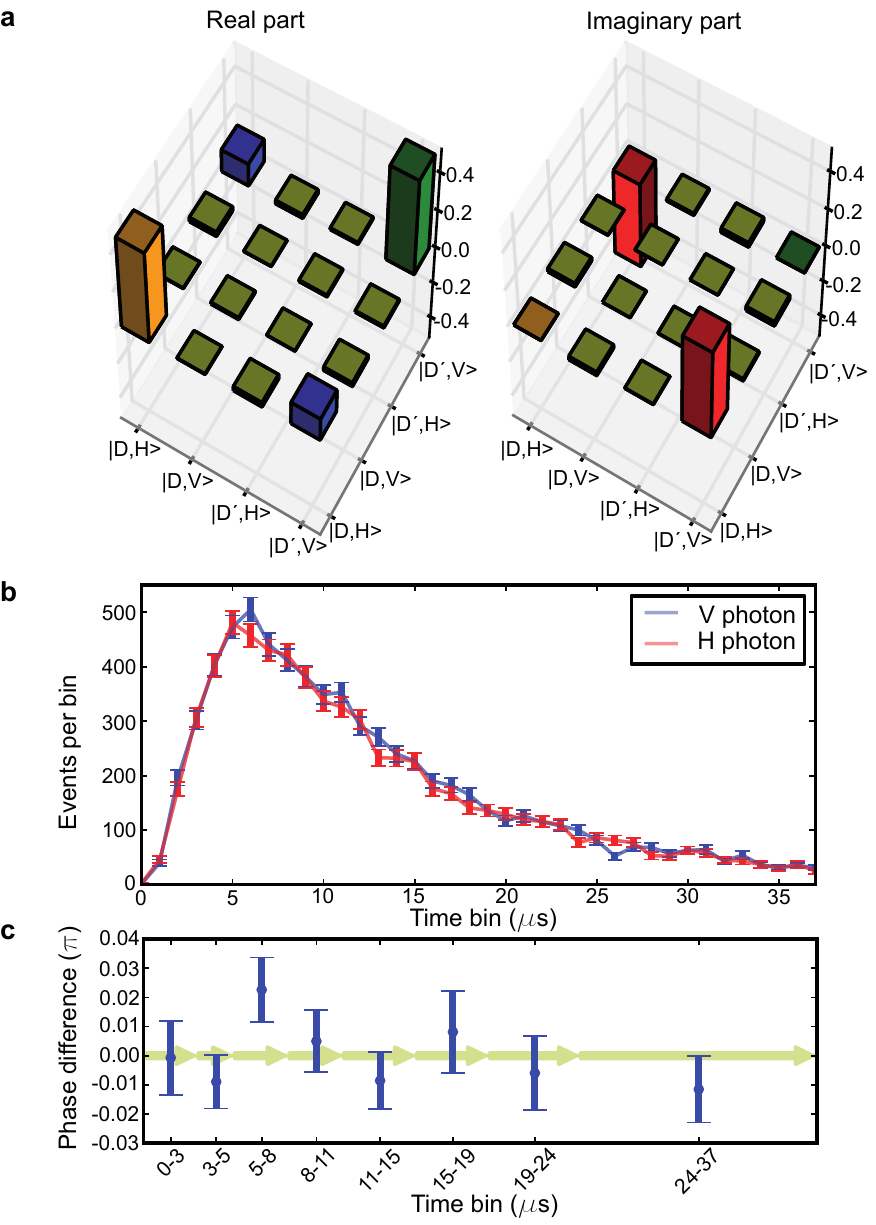}%
\caption{\label{photons} \textbf{Quantum state tomography of the joint ion-photon state, containing $\sim 40,000$ events.} \textbf{a}, Real and imaginary parts of all density matrix elements for Raman phase $\varphi=0.25$, from which a fidelity $F =$ \bestfidelity~is calculated.  Colors for the density matrix elements correspond to those used in Figs. 3a and 4a.  \textbf{b}, Temporal pulse shape of $H$ and $V$ cavity photons. Error bars represent one s.d. based on Poissonian photon statistics. \textbf{c}, Phase of the ion-photon state vs. photon-detection time.  Arrows indicate time-bin intervals of the tomography data. Error bars represent one s.d. (see Methods).}\end{figure}

In order to measure the ion in all three bases, we first
map the superposition of $\{D',D\}$ onto the $\{S,D\}$ states \cite{Leibfried03,Haeffner08}. We then perform additional coherent operations to select the measurement basis and discriminate between $S$ and $D$ via fluorescence detection \cite{Haeffner08}.
Each sequence lasts 1.5 ms and consists of 800~$\mu$s of Doppler cooling, 60~$\mu$s of optical pumping, a 40~$\mu$s Raman pulse, an 4~$\mu$s mapping pulse, an optional 4.3~$\mu$s rotation, and 500~$\mu$s of fluorescence detection.  The probability to detect a photon in a single sequence is \efficiency; we thus detect on average \rate~events/s. Note that the photon is generated with near-unit efficiency, and detection is primarily limited by the probability for the photon to exit the cavity (16\%) and the photodiode efficiencies (40\%).

In a first set of measurements, we choose the case $\alpha=\pi/4$, corresponding to a maximally entangled state $\ket{\psi}$. From the tomographic data, the density matrix is reconstructed as shown in Fig. \ref{photons}(a).
Here we have tuned $\Omega_{1}$ and $\Omega_{2}$ so as to produce both photon polarizations with equal probability, corresponding to maximal overlap of the temporal pulse shapes of $H$ and $V$ photons (Fig. (b)).
In order to demonstrate that the photon-detection time does not determine the phase of the state, we extract this phase from state tomography as a function of the photon time bin (Fig. \ref{photons}(c)). Because the frequency difference of the bichromatic fields $\Delta_1 - \Delta_2$ is equal to the level spacing between  $\ket{D}$ and $\ket{D'}$, the phase $\varphi=0.25 \pi$ remains constant. Further details are given in the Methods section.

Tomography over all time bins yields a fidelity of $F \equiv \bra{\psi}\rho\ket{\psi}$ = \bestfidelity~with respect to the maximally entangled state, placing our system definitively in the nonclassical regime $F > 50\%$. Another two-qubit entanglement witness is the concurrence \cite{Wootters98}, which we calculate to be \concurrence.
The observed entanglement can also be used to test local hidden-variable models (LHVMs) via the violation of the Clauser-Horne-Shimony-Holt (CHSH) Bell inequality \cite{CHSH69}.
Entanglement of a hybrid atom-photon system holds particular interest since it could be used for a loophole-free test of a Bell-type inequality \cite{rosenfeld09}.
While LHVMs require the Bell observable of the CHSH-inequality to be less than 2, we measure a value \chsh$>2$, where quantum mechanics provides an upper bound of $2\sqrt{2}$.

\begin{figure}
\begin{center}
\includegraphics[width=\columnwidth]{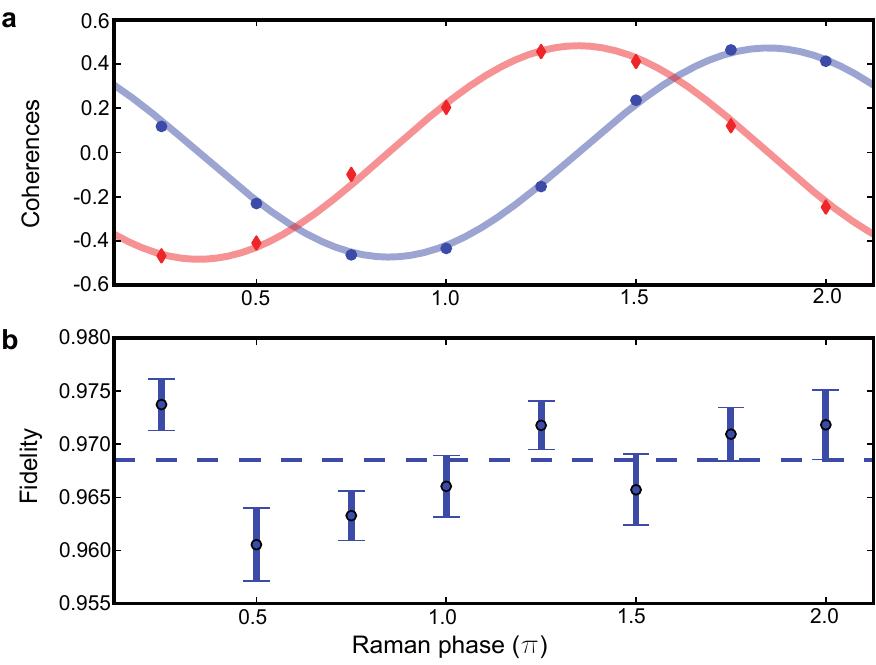}%
\caption{\label{tomography} \textbf{State tomography as a function of Raman phase ($\sim340,000$ events).} \textbf{a},
Re$(\rho_ {14})$ (blue circles) and Im$(\rho_ {14})$ (red diamonds)
as a function of Raman phase. Errorbars are smaller than the size of the symbols. Each value is extracted from a full state tomography of $\rho$ as in Fig \ref{photons}a. Both curves are fitted simultaneously, with phase offset constrained to $\pi/2$. The fit contrast is $95.6 (4) \% $. \textbf{b}, Fidelities of the eight states, with a dashed line indicating the mean value. Error bars represent one s.d. (see Methods).}
\end{center}
\end{figure}

We now establish that we can prepare $\ket{\psi}$ with high fidelity over the full range of the Raman phase $\varphi$.  We repeat state tomography for seven additional values of the relative Raman phase.
As a function of $\varphi$, the real and imaginary parts of the coherence $\rho_{14} \equiv \bra{DH}\rho\ket{D'V}$ vary sinusoidally as expected (Fig. \ref{tomography}(a)).
The fidelity has a mean value of \meanfidelity~and does not vary within error bars over all target phases (Fig. \ref{tomography}(c)).

\begin{figure}
\begin{center}
\includegraphics[width=\columnwidth]{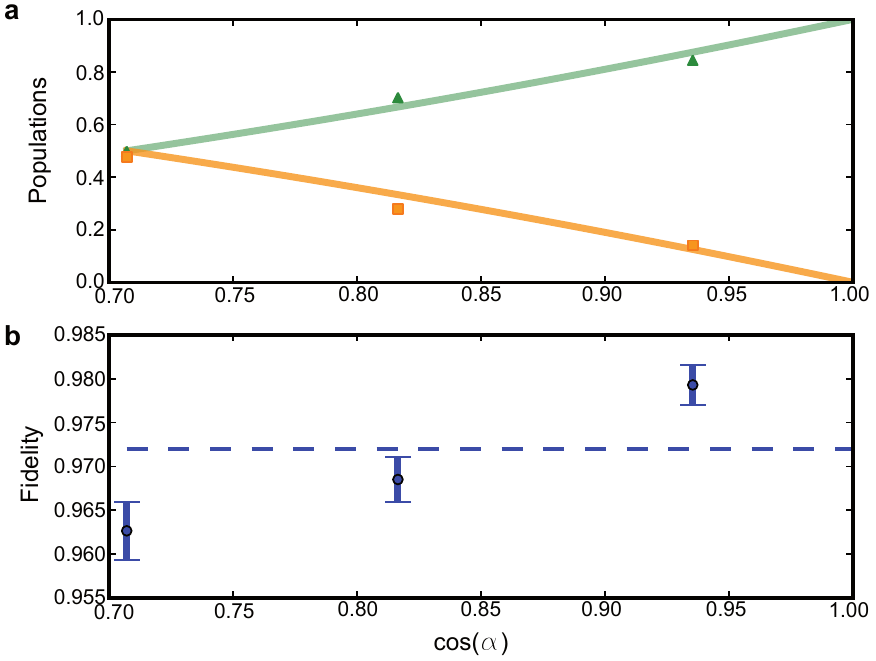}%
\caption{\label{raman} \textbf{State tomography for three values of amplitude $\cos{\alpha}$.} \textbf{a}, The density matrix elements $\rho_{11}$ (orange squares) and $\rho_{44}$ (green triangles) are plotted for the three target amplitudes  $\cos{\alpha}=\{ 1/\sqrt{2}, 1/\sqrt{3}, 1/\sqrt{8} \}$. Errorbars are smaller than the size of the symbols. Solid lines represent the amplitudes of the target states. \textbf{b}, The corresponding fidelities are $F=\{96.3 (3),~96.8 (3),~98.0(4)\}$.  A dashed line indicates the mean value. Error bars represent one s.d. (see Methods).}
\end{center}
\end{figure}

A second measurement set demonstrates control over the amplitudes $\cos{\alpha}$ and $\sin{\alpha}$ of the entangled ion-photon state. After selecting three target amplitudes $\cos{\alpha}=\{1/\sqrt{2},1/\sqrt{3},1/\sqrt{8}\}$,  we generate each corresponding state by adjusting the Raman field amplitudes, since $\alpha$ is a function of the ratio $\Omega_{2}/\Omega_{1}$.  The density matrix for each state is then measured.
In Fig. \ref{raman}(a), we see that the populations $\rho_{11} \equiv \bra{DH}\rho\ket{DH}$ and $\rho_{44} \equiv \bra{D'V}\rho\ket{D'V}$ for the three target amplitudes agree well with theoretical values.  The fidelities of the asymmetric states (Fig. \ref{raman}(b)) are as high as those of the maximally entangled states and are limited by the populations, that is, by errors in tuning the Raman fields to match the target values.

Errors in atomic state detection \cite{Volz06, rosenfeld09}, atomic decoherence \cite{Weber09} and multiple excitations of the atom \cite{Blinov04} reduce the fidelity of the atom-photon entangled state by $\ll 1 \%$. Imperfect initialization and manipulation of the ion due to its finite temperature and laser intensity fluctuations decrease the fidelity by 1\%.
The two most significant reductions in fidelity are due to dark counts of the APDs at a rate of 36~Hz (1.5\%) and imperfect overlap of the temporal pulse shapes (1\%).

To our knowledge, this measurement represents both the highest fidelity and the fastest rate of entanglement detection to date between a photon and a single-emitter quantum memory. This detection rate is limited by the fact that most cavity photons are absorbed or scattered by the mirror coatings, and only 16\% enter the output mode.  However, using mirrors with state-of-the-art losses and a highly asymmetric transmission ratio, an output  coupling efficiency exceeding 99\% is possible (see Methods). In contrast, without a cavity, using a lens of numerical aperture 0.5 to collect photons, the efficiency would be 6.7\%. In addition, the infrared wavelength of the output photons is well-suited to fiber distribution, enabling long distance quantum networks.  We note that a faster detection rate could be achieved by triggering ion-state readout on the detection of a photon.

We have demonstrated full control of the phase and amplitude of an entangled ion-photon state, which opens up new possibilities for quantum communication schemes. In contrast to monochromatic schemes, evolution of the relative phase of the atomic state after photon detection
is determined only by the start time of the experiment and not by the photon-detection time. The state $\ket{\psi}$ is in this sense predetermined and can be stored in, or extracted from, a quantum memory in a time-independent manner. The bichromatic Raman process employed here provides a basis for a coherent atom-photon state mapping as well as one- or two-dimensional cluster state generation \cite{Economou2010}.

We thank J. Barreiro, D. Nigg, K. Hammerer, and W. Rosenfeld for helpful discussions.
This work was supported by the Austrian Science Fund (FWF), the European Commission (AQUTE), the Institut f\"ur Quanteninformation GmbH,
and a Marie Curie International Incoming Fellowship within the 7th European Framework Program.

Experiments were performed by A.S., B.C., and T.E.N., with contributions from P.S. to the setup.  Data analysis was performed by A.S., B.C., and T.M.  The experiment was conceived by P.O.S. and R.B. and further developed in discussions with A.S., B.B., B.C., and T.E.N.  All authors contributed to the discussion of results and participated in manuscript preparation.

\section*{Methods}
\subsection{Detection and state tomography}
The cavity output path branches at a polarizing beamsplitter into two measurement paths, and the detection efficiencies of these paths are unequal. We compensate for this imbalance by performing two measurements for a given choice of ion and photon basis and sum the results; between the measurements, a rotation of the output waveplates swaps the two paths.

At each measurement setting, we record on average \correlations ~events in which a single photon has been detected. While a photon is detected in 5.7\% of sequences, the atom is always measured. Correlations of the photon polarization and the atomic state are the input for maximum likelihood reconstruction of the most likely states\cite{Jezek03}.
Error bars are one standard deviation derived from non-parametric bootstrapping\cite{Davison97} assuming a multinomial distribution.

\subsection{Time independence}
The phase of the entangled atom-photon state is inferred from the measurements of photon polarization and atomic-state phase. In the experiments of Refs.\cite{Blinov04, Wilk07b, Togan10}, although the phase of the entangled state is time independent before photon detection, the phase of the atomic state after photon detection evolves due to Larmor precession.
It is thus necessary to fix the time between photon detection and atomic state readout in order to measure the same $\varphi$ for all realizations of the experiment.
In contrast, for the case of Raman fields $\Omega_1 e^{i \omega_{1} t}$ and $\Omega_2 e^{i \omega_{2} t}$, the correct choice of frequency $\omega_{1}-\omega_{2} = \omega_{D'}-\omega_{D}$ means that both the phase of the entangled atom-photon state before photon detection and the phase of the atomic state after photon detection are independent of photon-detection time.

We define a model system with bases $\{ \ket{S,n}, \ket{D,n}, \ket{D',n} \}$, where $ n=\{0,1\}$ is the photon number in either of the two degenerate cavity modes.  The excited state has been adiabatically eliminated, so that $g^{\mathrm{eff}}_1$ couples $\ket{S,0}$ to $\ket{D,1}$ and $g^{\mathrm{eff}}_2$ couples $\ket{S,0}$ to $\ket{D',1}$.  After transformation into a rotating frame $U=e^{i\omega_{1}t\ket{S}\bra{S}}e^{i(\omega_{1}-\omega_{2})t\ket{D'}\bra{D'}}$, the Hamiltonian is
\begin{align*}
&(\omega_S-\omega_{1}) \ket{S}\bra{S} + \omega_D \ket{D}\bra{D} \\
&+ (\omega_{D'}-(\omega_{1}-\omega_{2})) \ket{D'}\bra{D'} + \omega_C  \ket{1}\bra{1} \\
&+ \left(g^{\mathrm{eff}}_1 \ket{D,1}\bra{S,0} + g^{\mathrm{eff}}_2 \ket{D',1}\bra{S,0}
+ \mathrm{h.c.} \right) ,
\end{align*}
where $\hbar=1$, $\{ \omega_S, \omega_D,  \omega_D' \}$ are the state frequencies, $ \omega_C$ is the cavity frequency, and terms rotating at $|\omega_{1}-\omega_{2}| \gg g^{\mathrm{eff}}_i$ are omitted\cite{shore1990theory}.
In this frame, the couplings $g^{\mathrm{eff}}_i$ are time-independent, and the states $\ket{D}$ and $\ket{D'}$ are degenerate. Therefore, the phase between $\ket{D,1}$ and $\ket{D',1}$ remains fixed during Raman transfer, and the phase between $\ket{D,0}$ and $\ket{D',0}$ stays constant after photon detection.

\subsection{Cavity parameters}
The cavity mirrors have transmission $T_1 = 13$~ppm and  $T_2 = 1.3$~ppm, with combined losses of 68 ppm.  State-of-the-art combined losses at this wavelength are $L = 4$~ppm\cite{Rempe91}.  In our cavity, these losses would correspond to an output coupling efficiency of $T_1/(T_1+T_2+L) = 71$\%.  To improve this efficiency, an output mirror with higher transmission $T_1$ could be used; for example, $T_1 = 500$~ppm corresponds to an efficiency of 99\%.   The cavity decay rate $\kappa$ would also increase, but single-photon generation with near-unit efficiency is valid in the bad-cavity regime\cite{Law97}.



\end{document}